# A new elliptical-beam method based on time-domain thermoreflectance (TDTR) to measure the in-plane anisotropic thermal conductivity and its comparison with the beam-offset method

Puqing Jiang, Xin Qian and Ronggui Yang[a]

*Department of Mechanical Engineering, University of Colorado, Boulder, Colorado 80309, USA*

Materials lacking in-plane symmetry are ubiquitous in a wide range of applications such as electronics, thermoelectrics, and high-temperature superconductors, in all of which the thermal properties of the materials play a critical part. However, very few experimental techniques can be used to measure in-plane anisotropic thermal conductivity. A beam-offset method based on time-domain thermoreflectance (TDTR) was previously proposed to measure in-plane anisotropic thermal conductivity. However, a detailed analysis of the beam-offset method is still lacking. Our analysis shows that uncertainties can be large if the laser spot size or the modulation frequency is not properly chosen. Here we propose an alternative approach based on TDTR to measure in-plane anisotropic thermal conductivity using a highly elliptical pump (heating) beam. The highly elliptical pump beam induces a quasi-one-dimensional temperature profile on the sample surface that has a fast decay along the short axis of the pump beam. The detected TDTR signal is exclusively sensitive to the in-plane thermal conductivity along the short axis of the elliptical beam. By conducting TDTR measurements as a function of delay time with the rotation of the elliptical pump beam to different orientations, the in-plane thermal conductivity tensor of the sample can be determined. In this work, we first conduct detailed signal sensitivity analyses for both techniques and provide guidelines in determining the optimal experimental conditions. We then compare the two techniques under their optimal experimental conditions by measuring the in-plane thermal conductivity tensor of a ZnO [11-20] sample. The accuracy and limitations of both methods are discussed.

---

[a] Electronic mail: Ronggui.Yang@Colorado.Edu

## I. INTRODUCTION

Materials lacking in-plane symmetry are ubiquitous in a wide range of applications such as electronics,[1,2] thermoelectrics,[3,4] high-temperature superconductors,[5,6] and thermal management,[7,8] in all of which the thermal properties play a critical part. However, very few of the thermal conductivity measurement techniques[9] can be used to measure in-plane anisotropic thermal conductivity. The beam-offset method based on the time-domain thermoreflectance (TDTR)[10,11] and the multiple-heater-line method based on the 3-omega[12,13] are both being developed over the recent years to measure the in-plane anisotropic thermal conductivity of small-scale (thin film) samples.

In the beam-offset method based on TDTR,[11] the pump beam is swept across the probe beam and the full-width half-maximum (FWHM) of the out-of-phase signal $V_{out}$ at a negative delay time of around -100 ps is used to derive the in-plane thermal conductivity along the scanning direction. One obvious disadvantage of the beam-offset method is that the derivation of the thermal properties relies solely on one data point, *i.e.*, the FWHM of $V_{out}$ at a negative delay time, whereas the uncertainty of the FWHM signal can have a significant impact on the measurement uncertainty. Besides, despite the initial demonstration of this method by Feser *et al.*,[11] it remains unclear how to choose the optimal experimental conditions that can yield the smallest measurement uncertainty.

In this paper, we propose an alternative approach using a highly elliptical pump beam based on TDTR to measure the in-plane thermal conductivity of laterally anisotropic materials. We first determine the optimal experimental conditions of the laser spot size and modulation frequency for both methods through detailed sensitivity analyses. We then discuss the accuracy and limitations of both methods in measuring the in-plane anisotropic thermal conductivity. In the end, the two methods are compared by measuring the in-plane thermal conductivity tensor of a ZnO [11-20] sample.

## II. METHODOLOGIES

Both the elliptical-beam method and the beam-offset method are based on TDTR, which is a powerful and versatile technique that has been applied to measure thermal properties of a wide range of thin films,[14-16] multilayers,[17,18] nanostructured and bulk materials,[19,20] and their interfaces.[21-23] TDTR uses two synchronized light sources, referred to as the pump (heating) and the probe (sensing) beams. The pump beam deposits a periodic heat flux on the sample surface and induces a temperature change in the sample, which is then monitored by measuring the change in the intensity of the reflected probe beam. A schematic diagram of a typical TDTR system is shown in Figure 1(a). More details of the system implementation have been described elsewhere.[24-29] Particularly, there are two features of the system relevant to this work that are worth mentioning here: (1) The polarizing beam splitter (PBS) in front of the objective lens is gimbal-mounted so that the pump beam can be steered to enable the operation of the beam-offset method while the position of the probe beam is unaffected. (2) A pair of cylindrical lenses can be added in the pump path to generate a highly elliptical pump beam for the elliptical-beam experiments. This pair of cylindrical lenses can be conveniently added or removed to change the shape of the pump beam if they are mounted on a magnetic kinematic base. In what follows, we will have detailed discussions on the currently proposed elliptical-beam method and the recently developed beam-offset method,[11] respectively. We will specifically focus on their optimal experimental conditions, measurement accuracy, and limitations.

## A. Elliptical-beam method

In our proposed elliptical-beam TDTR approach, the experiments are conducted following the same procedure as in the conventional TDTR, *i.e.*, the pump and the probe beams are concentrically aligned and the ratio signals $R = -V_{in}/V_{out}$ acquired as a function of delay time are used to derive the thermal properties, except that the pump beam is of a highly elliptical shape. A schematic of the elliptical-beam method is shown in Figure 1(b), the rationale of which is described in details below.

In TDTR experiments, the heat energy is first deposited by the pump beam on the sample surface, and then it spreads out in all directions, as illustrated in Figure 1(c). Due to the multiple timescales in TDTR experiments, the heat source includes two parts: one is the pulsed heating and the other is the sinusoidal, *continuous* heating at the modulation frequency. The pulsed heating mainly induces a jump in the in-phase temperature response at the zero delay time and its subsequent cooling before the coming of the next pulse: $\Delta V_{in} = V_{in} - V_{in}(t_d < 0)$, where $V_{in}(t_d < 0)$ is the so-called pulse accumulation.[30] On the other hand, the modulated continuous heating mainly induces the out-of-phase temperature response $V_{out}$.[31] Therefore, the in-phase signal $\Delta V_{in}$ that is induced by single pulse heating depends on the delay time but not on the modulation frequency, while the out-of-phase signal $V_{out}$ that is induced by modulated continuous heating depends on the modulation frequency but not on the delay time. The thermal diffusion length of the pulsed heating can be estimated as $d_f = \sqrt{K\tau/C}$, where $K$ and $C$ are the thermal conductivity and volumetric heat capacity of the substrate, respectively, and $\tau$ is the relaxation time of the in-phase temperature response after the pulse heating. In a typical TDTR system using an 80 MHz repetition rate pulsed laser, the maximum detectable relaxation time is limited by the time interval between the pulses, *i.e.*, $\tau \leq 12.5$ ns (=1/80 MHz). On the other hand, the heat spreading length scale of the modulated continuous heating can be represented by a thermal penetration depth, defined as $d_p = \sqrt{K/\pi f C}$, where $f$ is the modulation frequency. Generally, the thermal diffusion length of the pulse heating $d_f$ is one order of magnitude smaller than the thermal penetration depth of the continuous heating $d_p$ and is also much smaller than the laser spot size. Therefore, for most cases, the pulsed heating is mainly one-dimensional in the through-plane direction, with the in-phase temperature response $\Delta V_{in}$ being sensitive to the thermal properties of the sample only in the through-plane direction. On the other hand, the continuous heating can be either one-dimensional or three-dimensional, depending on the comparison of the laser spot size with the in-plane thermal penetration depth $d_{p,in}$. When TDTR experiments are conducted using a laser spot size much larger than the in-plane thermal penetration depth $d_{p,in}$, the heat flow will be mainly one-dimensional and the out-of-phase signal $V_{out}$ will be mainly sensitive to the thermal properties of the sample in the through-plane direction. When TDTR experiments are conducted using a tightly focused laser spot with the size similar or even smaller than the in-plane thermal penetration depth $d_{p,in}$, the out-of-phase signal $V_{out}$ will be sensitive to the thermal properties of the sample in both the in-plane and through-plane directions. For samples lacking in-plane symmetry, we can thus selectively suppress the sensitivities of the out-of-phase signal $V_{out}$ to thermal properties along certain in-plane directions by using a highly elliptical laser spot for the TDTR measurements.

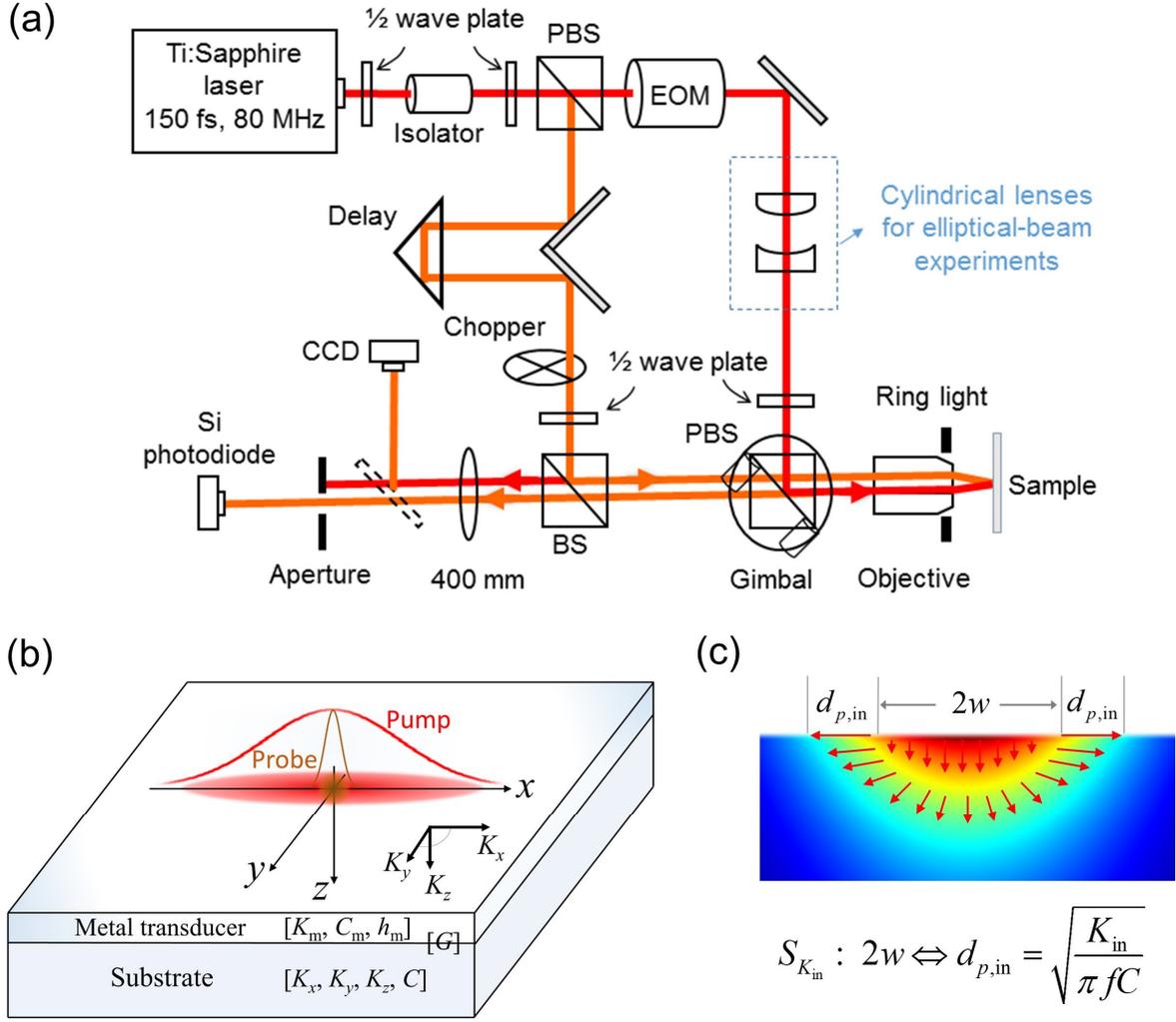

FIG. 1. (a) Schematic of the TDTR setup. The acronyms EOM, PBS, and BS stand for electro-optic modulator, polarizing beam splitter, and beam splitter, respectively. (b) Schematic of the elliptical-beam method for measuring in-plane anisotropic thermal conductivity. (c) Illustration of heat flux directions in TDTR experiments and how the laser spot size $2w$ as compared to the in-plane thermal penetration depth $d_{p,in}$ determines the sensitivity of the TDTR signals to the in-plane thermal conductivity $K_{in}$.

A criterion is thus needed for us to decide how the elliptical beam spot sizes should be chosen along the major and minor axes so that the detected TDTR signals is selectively sensitive to the in-plane thermal properties along the minor axis of the elliptical beam. How the detected signals are sensitive to the parameter $\alpha$ can be quantified by defining a sensitivity coefficient[29]

$$S_\alpha \equiv \frac{\partial \ln R}{\partial \ln \alpha} = \frac{\partial R/R}{\partial \alpha/\alpha} \tag{1}$$

which represents the relative change in the ratio signal $R = -V_{in}/V_{out}$ with respect to the relative change in the parameter $\alpha$. ($S_\alpha = 0$ means that the TDTR measurements are not affected by $\alpha$. With a larger amplitude of $S_\alpha$, TDTR measurements are more strongly dependent on $\alpha$.) To calculate the sensitivity coefficients, an accurate thermal transport model is needed to predict the detected signals in TDTR experiments. A detailed derivation of the thermal model for the elliptical-beam method is presented in Appendix A, following the model developed for circular beam spots.[32] Briefly, there are several input parameters in the thermal model, including the thermal conductivity, volumetric heat capacity, and thickness of the transducer layer ($K_m$, $C_m$, $h_m$), the thermal conductivity values of the substrate along different directions ($K_x$, $K_y$, $K_z$), volumetric heat capacity of the substrate ($C$), interface thermal conductance between the transducer and the substrate ($G$), and the $1/e^2$ laser spot radii along the major and the minor axis ($w_x$, $w_y$), see an illustration of the sample configuration in Figure 1(b). Note that in this text the symbol $w_x$ is used to represent the root mean square (RMS) average of the $1/e^2$ radii of the pump $\sigma_{0x}$ and the probe $\sigma_{1x}$ in the x-direction $w_x = \sqrt{(\sigma_{0x}^2 + \sigma_{1x}^2)/2}$, and likewise for $w_y$. We focus only on $w_x$ and $w_y$ because the TDTR signals depend on the RMS average of the pump and probe sizes rather than their individual sizes.

Figure 2 (a-c) shows the sensitivity coefficients of the ratio signals $R$ in elliptical-beam experiments to the in-plane thermal conductivity $K_x$ of quartz, ZnO, and graphite as a function of the laser spot size $w_x$ and in-plane thermal penetration depth $d_{p,x}$. Here the x-direction is chosen as the representative one to discuss how the laser spot size $w_x$ as compared to the in-plane thermal penetration depth $d_{p,x}$ along the same direction affects the sensitivity to $K_x$, while the discussion here also applies to any other in-plane direction. The transducer layer is chosen as the 100-nm-thick Al film, and the delay time is fixed to be 100 ps. The three samples of quartz, ZnO, and graphite are taken as the representative ones for the sensitivity analysis here because they have a wide range of in-plane thermal conductivity spanning from 10 W m$^{-1}$ K$^{-1}$ to 2000 W m$^{-1}$ K$^{-1}$ and a wide range of anisotropic ratio $K_x/K_z$ from 1.5 to 300. Figure 2 shows that the sensitivity plots of the three samples all exhibit the same trend that the ratio signal is highly sensitive to $K_x$ in the limit $d_{p,x} \gg w_x$ and not sensitive to $K_x$ in the other limit $d_{p,x} \ll w_x$. Following the same approach in Ref. 33, we set the thresholds that the sensitivity $S_\alpha > 0.2$ is considered "high" and the sensitivity $S_\alpha < 0.05$ is considered "low". In other words, a sensitivity $S_{K_x} > 0.2$ is needed for the signal to be considered sufficiently sensitive to $K_x$ and a sensitivity $S_{K_x} < 0.05$ is needed for the signal to be considered not sensitive to $K_x$. From the sensitivity plots in Figure 2, we find that the criterion $w_x < 2d_{p,x}$ generally guarantees a high sensitivity to $K_x$ except for the quartz sample with Al transducer. The reason is that with the low thermal conductivity of the substrate (quartz) in contrast to the high thermal conductivity of the transducer layer (Al film), the majority of the laser heat spreads out in the transducer layer instead of penetrating to the substrate, making the ratio signal less sensitive to the in-plane thermal conductivity of the substrate. Therefore, a more stringent criterion of $w_x < d_{p,x}$ would be needed for low-thermal-conductivity materials such as quartz to have a sufficient sensitivity to $K_x$. On the other hand, the criterion $w_x > 5d_{p,x}$ guarantees that the sensitivities to $K_x$ are effectively suppressed for all the samples.

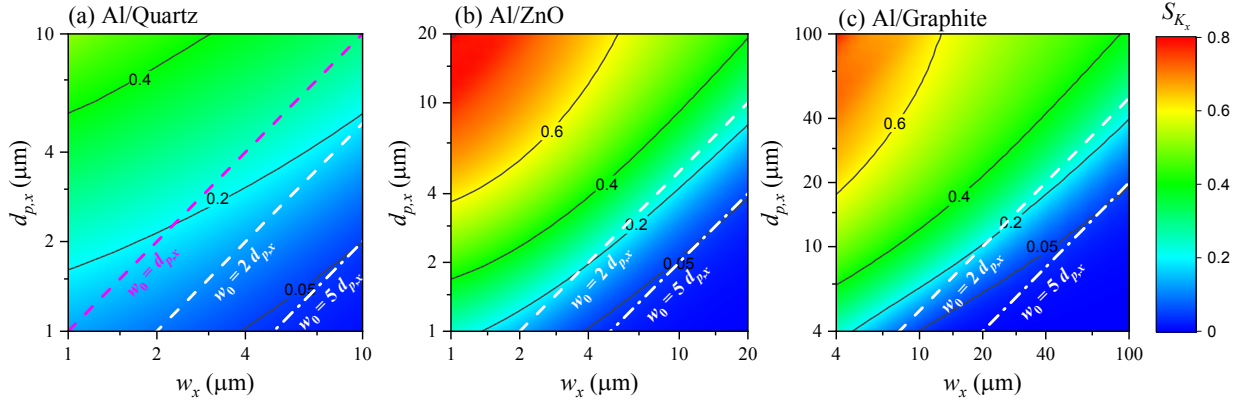

FIG. 2. Sensitivity coefficients of the ratio signal in elliptical-beam experiments to $K_x$ of the substrate as a function of laser spot size $w_x$ and in-plane thermal penetration depth $d_{p,x}$ along the same direction for different samples of quartz (a), ZnO (b), and graphite (c), with 100 nm Al as the transducer. The delay time is fixed to be 100 ps.

Figure 3(a) shows an example of the sensitivity coefficients of the ratio signal in elliptical-beam experiments to the different parameters of a ZnO sample, with a 100-nm-thick Al film as the transducer. The elliptical laser spot has the sizes of $w_x = 4$ μm and $w_y = 20$ μm, and the modulation frequency is set at 0.35 MHz. The in-plane thermal penetration depth in ZnO at 0.35 MHz is estimated to be ~4 μm; so the choices of $w_x$ and $w_y$ here should meet the criteria to suppress the sensitivity to $K_y$ and maintain the sensitivity to $K_x$. We can see that the signals are sensitive to both $K_x$ and $K_z$ but in different manners as a function of delay time. While the sensitivity to $K_x$ is constant over the whole delay time range of 0.1 – 10 ns, the signal is sensitive to $K_z$ only in the short delay time range but not in the long delay time ranger > 2 ns. A more detailed analysis on the sources of sensitivity to $K_x$ and $K_z$ and an explanation on their uncorrelated behavior as a function of delay time can be found in Appendix B. We can thus simultaneously determine these parameters from one single set of the elliptical-beam measurements using a least squares algorithm developed by Yang et al.[34] and widely used by others for TDTR experiments.[19,20] Figure 3(a) shows that the dominant sources of uncertainty for the elliptical-beam experiments come from the minor radius of the elliptical beam $w_x$, the thickness of the transducer film $h_m$, and the heat capacity of the transducer film $C_m$. Among the input parameters, we assume an uncertainty of 10% for $K_m$, $K_y$, and $G$, 3% for $C_m$ and $C_{sub}$, 4% for $h_m$, and 3% for $w_x$ and $w_y$ for the uncertainty estimation. Figure 3(b) shows the confidence range of $K_x$ and $K_z$ when these two parameters are determined simultaneously from the elliptical-beam experiment. In this case, $K_x$ and $K_z$ of the substrate can be determined with an uncertainty of 8.4% and 16.1%, respectively. As we will see later that a significant advantage of the elliptical-beam method over the beam-offset method is that multiple parameters can be determined simultaneously from one single set of the measurements, while in the beam-offset method there is only one data point, making the input parameters 100% correlated to each other.

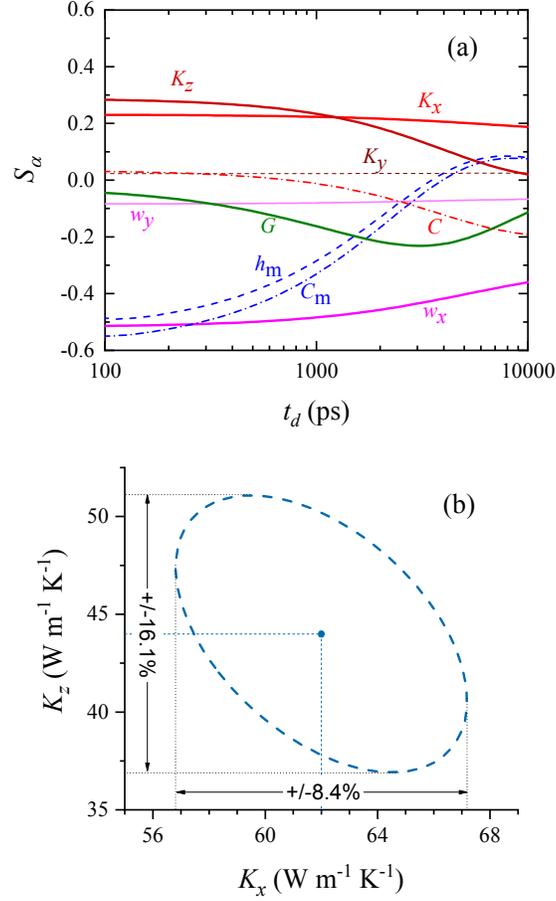

FIG. 3. (a) Sensitivities of the ratio signal in the elliptical-beam method to different parameters of the ZnO sample, plotted as a function of delay time. The elliptical laser spot has the sizes of $w_x$ = 4 μm and $w_y$ = 20 μm, and the modulation frequency is 0.35 MHz. The transducer layer is 100 nm Al. (b) Confidence ranges of $K_x$ and $K_z$ of ZnO with these two parameters simultaneously determined from the elliptical-beam experiment. The uncertainties of $K_x$ and $K_z$ are estimated as 8.4% and 16.1%, respectively.

## B. Beam-offset method

In the beam-offset TDTR experiments, the signals were taken as the FWHM of the out-of-phase signal $V_{out}$ at a negative delay time, e.g., $t_d$ = -100 ps. There are several reasons for this practice:[11] (1) The FWHM of $V_{out}$ is almost exclusively sensitive to the thermal conductivity along the offset direction but not to the thermal conductivity along the orthogonal directions. (2) The FWHM is independent of the absolute amplitude of the temperature fluctuation when the steady-state temperature rise is <10 K. (3) The amplitude of $V_{in}$ at a negative delay time is relatively small so that it alters $V_{out}$ less for any small error in the reference phase of the lock-in amplifier. In addition, Feser et al.[11] recommended using a thin NbV film instead of the conventional 100-nm-thick Al film as the metal transducer for the beam-offset TDTR experiments. The reason is that with a much-reduced thermal conductance of the metal film ($K_m h_m$), the FWHM signal will be more sensitive to the thermal conductivity of the substrate and less sensitive to the properties of the transducer layer, thus yielding a smaller measurement uncertainty.[11]

However, despite the initial demonstration of this technique by Feser et al.[11], it is still unclear to the readers how the laser spot size and modulation frequency should be chosen and whether the NbV transducer is always a necessary replacement of the conventional Al transducer for the beam-offset experiments. To find out the optimal configurations of the laser spot size $w_0$ and modulation frequency $f$ for the beam-offset experiments, we calculate the sensitivities of the FWHM signal to $K_x$ of the substrate for quartz, ZnO, and graphite with different metal transducers of NbV and Al as a function of $w_0$ and $f$. Here the symbol $w_0$ is used because the laser spot is generally circular with $w_x = w_y = w_0$ in the beam-offset experiments. The FWHM signals in the beam-offset experiments are calculated using the thermal transport model outlined in Appendix A. Note that the same definition of the sensitivity coefficient as in Eq. (1) also applies to the beam-offset method, only to have the ratio signal $R$ replaced with the FWHM signal in the beam-offset experiment. For the analysis here, the offset direction is assumed to be parallel to the $x$-coordinate so the FWHM signal is sensitive to $K_x$ of the substrate. We find that when the frequency is converted into a length scale, the in-plane thermal penetration depth along the $x$-direction $d_{p,x}$, the sensitivity $S_{K_x}$ has peak values when $w_0 \approx d_{p,x}$ despite the wide range of in-plane thermal conductivity and anisotropic ratio of the three samples and the different metal transducers, as shown in Figure 4 (a-f).

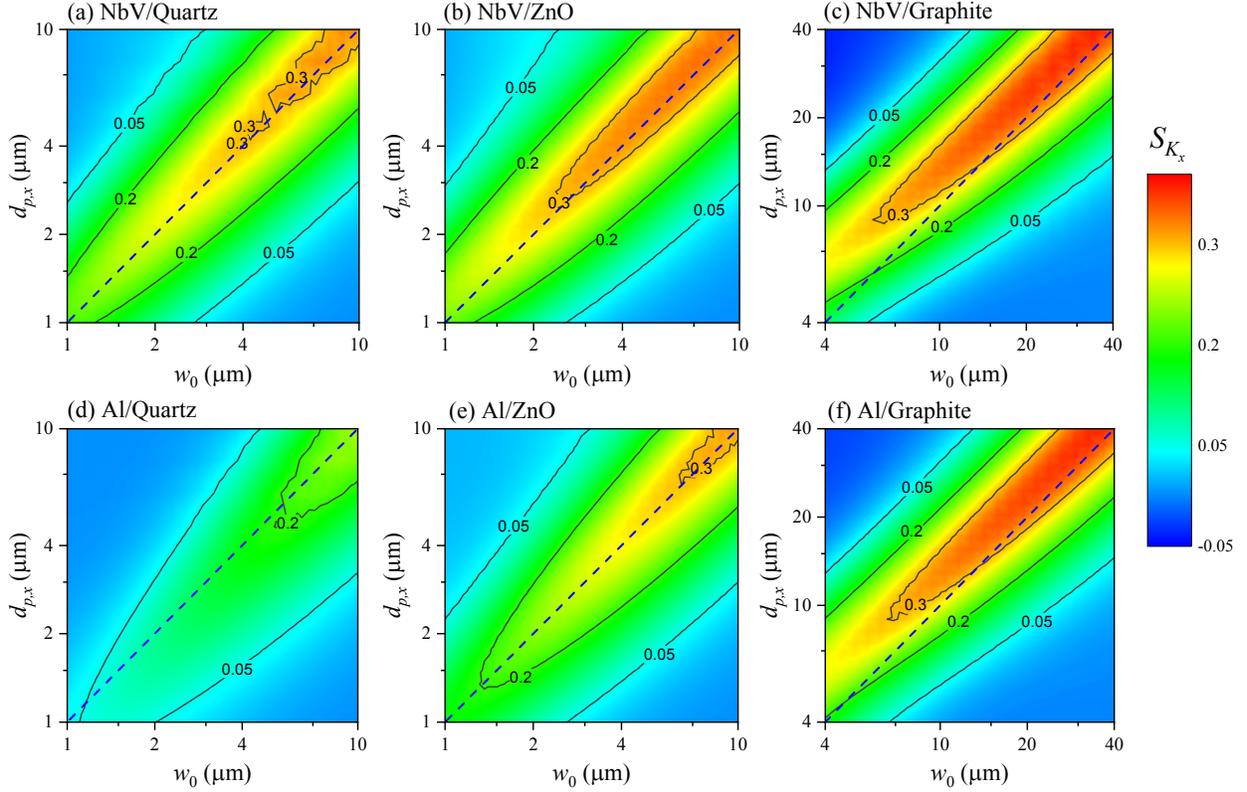

*FIG. 4. Sensitivity coefficients of the FWHM signal in beam-offset experiments to $K_x$ of the substrate as a function of laser spot size $w_0$ and in-plane thermal penetration depth $d_{p,x}$ for different samples of quartz, ZnO, and graphite, with different transducers of NbV and Al.*

The conclusion that the laser spot size $w_0$ being similar to the in-plane thermal penetration depth $d_{p,x}$ yields the highest sensitivity to $K_x$ sounds counter-intuitive at a first glance, as the out-of-phase signal $V_{out}$ is expected to be highly dependent on $d_{p,x}$ when $d_{p,x} \gg w_0$. To better understand the effect of $d_{p,x}$ on the FWHM signal, we plot out the simulated $V_{out}$ as a function of offset distance $x_c/w_0$ for the quartz sample at different modulation frequencies of 1 MHz and 0.1 MHz in Figure

5. The original $V_{out}$ signals are compared with those simulated with $K_x$ increased by 10% as a convenient demonstration of how the $V_{out}$ signals are affected by $K_x$. The comparison in Figure 5 shows that for the quartz sample measured at 1 MHz with $d_{p,x} \approx w_0$, only the short offset range $x_c < w_0$ of the $V_{out}$ signals are affected by $K_x$, while for the measurements at 0.1 MHz with $d_{p,x} \gg w_0$, the $V_{out}$ signals are affected by $K_x$ almost proportionally over the whole offset distance range. Therefore, with $d_{p,x} \gg w_0$, the FWHM signal remains almost unchanged despite the fact that the amplitude of the $V_{out}$ signal is significantly affected by $K_x$. On the other hand, when $d_{p,x} \ll w_0$, the $V_{out}$ signals do not depend on $K_x$ (not shown in Figure 5) and the FWHM signal is only sensitive to the laser spot size $w_0$. Since the FWHM signal is not sensitive to $K_x$ for both the limits of $d_{p,x} \gg w_0$ and $d_{p,x} \ll w_0$, we can conclude that the FWHM signal has the peak sensitivity to $K_x$ when $d_{p,x} \approx w_0$.

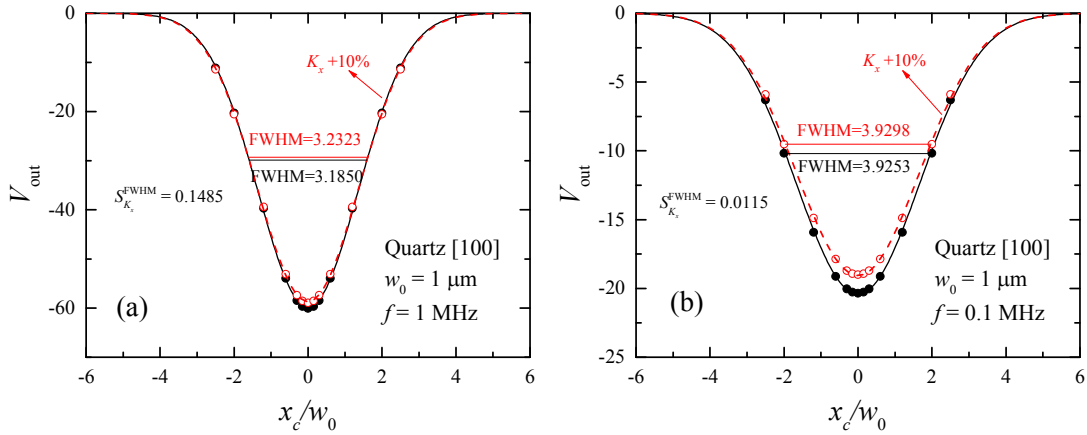

FIG. 5. *The simulated out-of-phase signal of the quartz sample as a function of beam offset distance at different modulation frequencies of (a) 1 MHz, and (b) 0.1 MHz. The symbols are the model simulations and the curves are the Gaussian fitting of the data points. The open symbols and the dashed curves are calculated with +10% increase in the nominal $K_x$ value to demonstrate how the $V_{out}$ signals are sensitive to $K_x$.*

Figure 4 also shows that the type of the transducer film significantly affects the sensitivity to $K_x$ only for the quartz sample but not for the high thermal conductivity samples such as ZnO or graphite. A more straightforward way to evaluate the effect of the transducer is to compare the error propagations for $K_x$ of the substrate estimated using the formula

$$\eta_{K_x} = \sqrt{\sum_\alpha \left(\frac{S_\alpha}{S_{K_x}} \eta_\alpha\right)^2} \qquad (2)$$

where $\eta$ is the uncertainty in percentage, and $\alpha$ is any input parameter except $K_x$. Among the input parameters, we assume an uncertainty of 10% for $K_m$, $K_y$, $K_z$, and $G$, 3% for $C_m$ and $C_{sub}$, 4% for $h_m$, and 3% for $w_0$. To calculate the uncertainties of $K_x$, the laser spot sizes were chosen as $w_0$ = 2 μm, 4 μm, and 10 μm for the quartz, ZnO, and graphite sample, and the appropriate $f$ is chosen for each sample so that $d_{p,x} = w_0$. The results are presented as the column bars in Figure 6, from which it is found that the NbV transducer in replacement of the conventional Al transducer dramatically reduces the measurement uncertainty from 35% to 7% for the quartz sample but makes a negligible difference for the ZnO or the graphite sample.

To have a better understaning on the effect of the metal transducer, the uncertainty $\eta_{K_x}$ is calculated for a series of hypothetical samples with a wide range of $K_x$. The hypothetical samples are assumed to have a constant heat capacity of $C$ = 2.0 MJ m$^{-3}$ K$^{-1}$ and an isotropic thermal conductivity $K_x = K_y = K_z$. Here the modulation frequency is fixed at $f$ = 1 MHz and the laser spot size is chosen as $w_0 = d_{p,x}$. From the results shown in Figure 6, NbV transducer is found to be very effective in reducing the measurement uncertainty when the substrate has its in-plane thermal conductivity in the range $K_x$ = 6 – 30 W m$^{-1}$ K$^{-1}$. For samples with $K_x$ > 30 W m$^{-1}$ K$^{-1}$, both NbV and Al transducers work well, giving an uncertainty of <15% for the $K_x$ measurements. On the other hand, for samples with $K_x < 6$ W m$^{-1}$ K$^{-1}$, neither NbV nor Al transducer work, as the measurement uncertainty becomes very high for both transducers. However, we should note that the uncertainty $\eta_{K_x}$ presented in Figure 6 is only the part propagated from the uncertainties of the input parameters. In practical experiments, the uncertainty of the FWHM signals could still contribute another significant part to the total uncertainty of $K_x$.

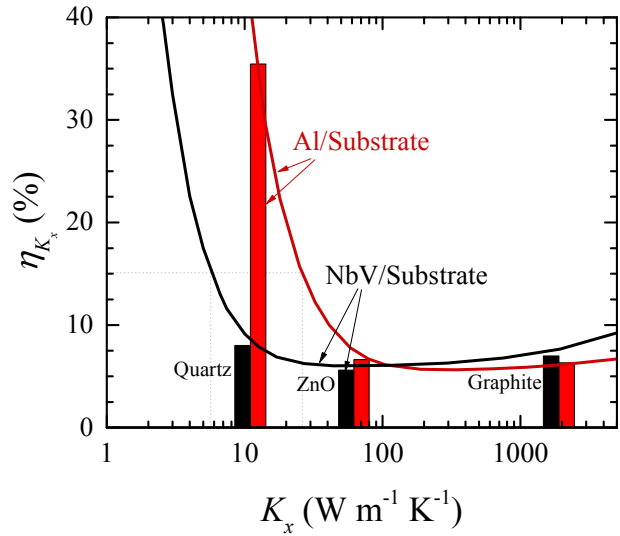

FIG. 6. Estimated uncertainty of measured $K_x$ as a function of $K_x$ of the substrate using Al and NbV transducers, respectively. Note that the uncertainty presented here only includes the error propagation from the input parameters but has not taken into account the uncertainty of the FWHM signals in real experiments.

## III. COMPARISON OF THE TWO METHODS VIA EXPERIMENTAL DEMONSTRATION

In this section, we compare the elliptical-beam method and the beam-offset method based on TDTR by measuring the in-plane thermal conductivity tensor of a ZnO [11-20] sample (optical grade, purchased from MTI Corp) using both techniques. ZnO is a hexagonal wurtzite crystal and has a higher thermal conductivity parallel to its *c*-axis (55 – 62 W m$^{-1}$ K$^{-1}$ from literature[29,35]) than the other directions (~44 W m$^{-1}$ K$^{-1}$ from literature[29,35]). In the ZnO wafer that is [11-20] oriented (a-plane), the *c*-axis lies in-plane, see the inset of Figure 10 for an illustration of the wafer orientation. To prepare the samples for TDTR measurements, a layer of 100 nm Al film on the samples was deposited as the transducer using e-beam evaporator. The Al layer thickness was determined by picosecond acoustics,[36] with an uncertainty of ~4%. The ZnO wafer was cleaned from any organic residue using isopropyl alcohol and ethanol before the metal deposition.

To conduct the elliptical-beam experiment, a pair of cylindrical lenses were inserted in the pump path to generate a highly elliptical pump laser spot on the focal plane of the objective lens, a schematic of which is shown in Figure 7(a). The cylindrical lenses only compress the pump beam in the horizontal direction so that the focused pump spot on the focal plane of the objective lens will be elongated in the horizontal direction but the spot size in the vertical direction is unaffected. The laser spot size was characterized by sweeping the pump spot across the probe spot and measuring the in-phase signal $V_{in}$ as a function of the offset distance $x_c$ at a high modulation frequency of 10 MHz and a short positive delay time of 100 ps. The laser spot size was extracted by fitting the $V_{in}$ profile to a Gaussian function $V_{in} \sim \exp(-x_c^2/w_0^2)$. The offset of the pump beam was controlled by rotating the polarizing beam splitter (PBS) that steers the pump beam. In this work, two program-controlled actuators are used to control the rotation of the PBS. The pump beam can thus be offset to the probe beam in any given direction with a minimum spatial resolution of 0.2 μm.

When using the elliptical-beam method to measure the in-plane thermal conductivity tensor, one can either rotate the cylindrical lens to control the shape of the pump beam, or fix the pump beam and rotate the sample. In practice, we choose the latter so that the laser spot size is fixed without the need to be characterized for each measurement. The sample is placed on an indexing mount that can be rotated for 360 degrees with an accuracy of 0.5 degree. On the other hand, when using the beam-offset method to measure the in-plane thermal conductivity tensor, the sample can be fixed while the beams are offset in any direction. Since the laser spot might not be 100% circular, the laser spot sizes are carefully characterized for each offset direction before conducting the beam-offset experiments. Figure 7(b) shows the characterized laser spot size in different directions for the beam-offset experiments. The laser spot has an average radius of $w_0 = 4.7$ μm, with the elliptical rate <5%. Figure 7(c) shows the averaged laser spot sizes of the pump and the probe along the major axis and the minor axis of the elliptical beam for the elliptical-beam experiments. The averaged elliptical beam has a long radius of 17.3 μm and a short radius of 4.5 μm. A low modulation frequency of 0.35 MHz is chosen for both the beam-offset experiment and the elliptical-beam experiment so that the in-plane thermal penetration depth $d_{p,x}$ and the laser spot sizes $w_x$ meet the optimal experimental conditions for both techniques.

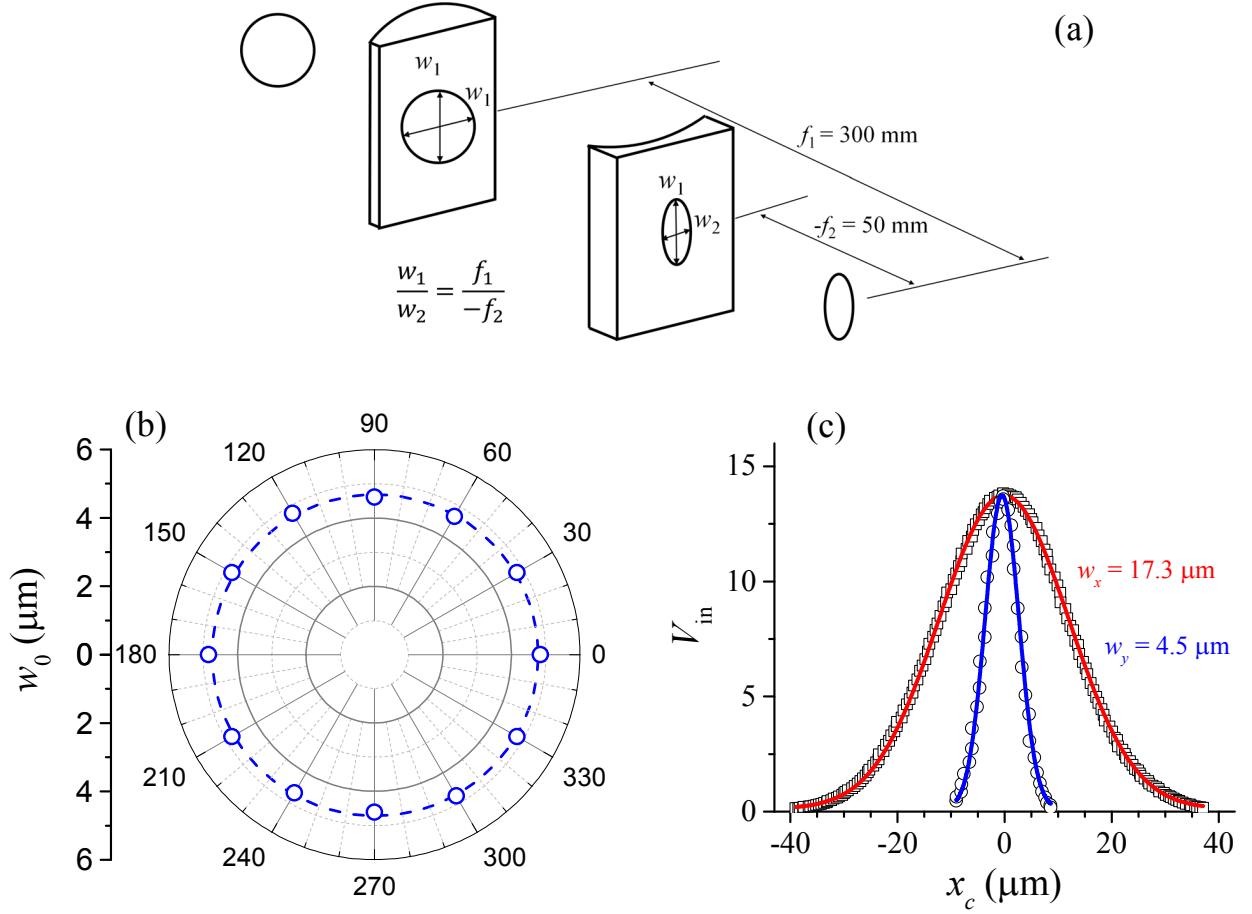

FIG. 7. *(a) Schematic of a pair of cylindrical lenses to generate an elliptical laser beam. (b) Circular laser spot sizes characterized for the beam-offset experiment along different offset directions. (c) Elliptical laser spot sizes (averaged between the pump spot and the probe spot) along the major and minor axes for the elliptical-beam experiment.*

Figure 8 (a) shows an example of the $V_{out}$ signal as a function of offset distance $x_c$ averaged from the signals of ten individual measurements, with the offset direction perpendicular to the *c*-axis of the sample. The FWHM is determined as 11.3 μm by fitting the $V_{out}$ signals using a Gaussian function. In order to determine $K_x$ from the measured FWHM, we use the thermal transport model outlined in Appendix A to simulate the FWHM as a function of $K_x$, as shown by the thick solid line in Figure 8 (b). $K_x$ of the substrate along the offset direction can thus be found by matching the measured FWHM to the simulated FWHM. We thus extract $K_a$ = 38.5 W m$^{-1}$ K$^{-1}$ along the direction perpendicular to the *c*-axis of ZnO [11-20].

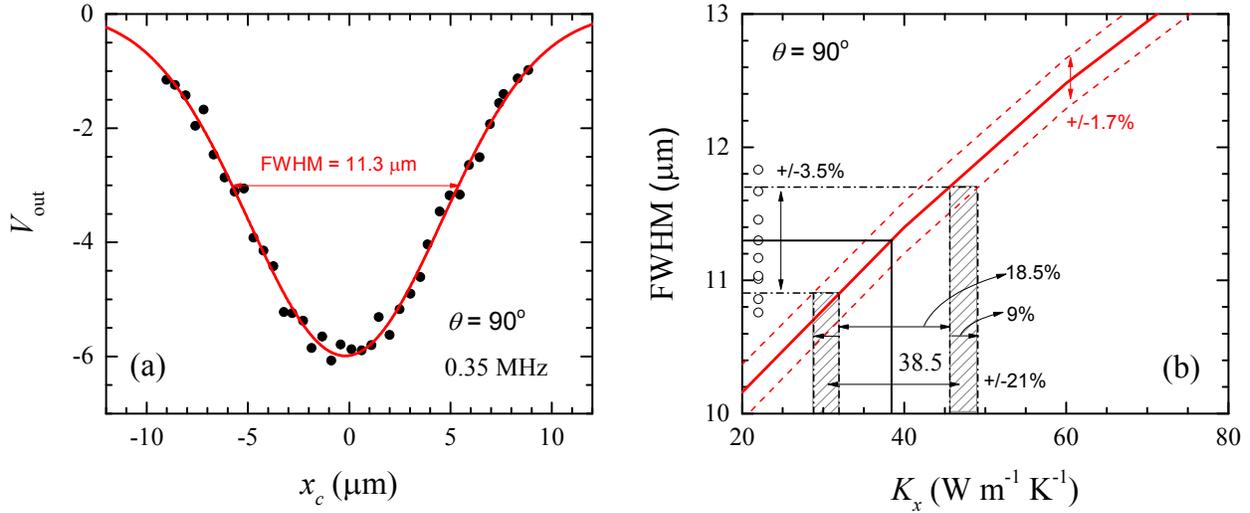

*FIG. 8. (a) Averaged out-of-phase signal $V_{out}$ as a function of offset distance from ten individual measurements (symbols) for the ZnO sample measured using modulation frequency 0.35 MHz, with the FWHM determined from the fitted Gaussian function (curves). (b) Determination of $K_x$ and its uncertainty by comparing the measured FWHM with the simulated FWHM.*

Figure 8(b) shows that there are two major sources of uncertainty for the determined $K_x$ in beam-offset experiments. One is the measured FWHM that has an uncertainty of repeatability due to the experimental noise and the error in determining the reference phase of the lock-in detection; the other is the simulated FWHM that has an error propagated from the uncertainties of the input parameters in the thermal transport model. To determine the repeatability of the measured FWHM, we individually repeated the measurements for ten times and found that the measured FWHM signal has an uncertainty of ±3.5% for our case. The uncertainty of the simulated FWHM is determined using the formula

$$\eta_{\text{FWHM}} = \sqrt{\sum_{\alpha}(S_\alpha \eta_\alpha)^2} \qquad (3)$$

where $\alpha$ is any input parameter except $K_x$. Assuming an uncertainty of 10% for $K_m$, $K_y$, $K_z$, and $G$, 3% for $C_m$ and $C_{\text{sub}}$, 4% for $h_m$, and 3% for $w_x$ and $w_y$, we estimate that the simulated FWHM has an uncertainty of ±1.7%. Figure 8 (b) shows that the ±3.5% uncertainty from the measured FWHM causes ±18.5% uncertainty in $K_x$, while the ±1.7% uncertainty in the simulated FWHM causes ±9% uncertainty in $K_x$. Since these two sources of uncertainty are independent of each other, the total uncertainty for the $K_x$ is determined as $\eta_{K_x} = \pm\sqrt{18.5^2 + 9^2}\% = \pm 21\%$.

Figure 9 (a) and (b) show the ratio signals from the elliptical-beam experiments and their fitting curves for ZnO [11-20] with the short radius of the elliptical beam parallel and perpendicular to the *c*-axis of ZnO, respectively. The curves of 30% bounds on the best-fitted thermal conductivity values are also included as a guide of reading to the sensitivity of the signals. It can be seen that the sensitivity to $K_x$ is successfully suppressed by the large $w_x$ of the elliptical beam while the sensitivity to $K_y$ is maintained due to the small $w_y$ in the *y*-direction. The sensitivity to $K_y$ is uniform over the whole delay time range, while the sensitivity to $K_z$ diminishes in the long delay time range > 2 ns. This means that $K_y$ and $K_z$ can be determined simultaneously from one set of the measurement. From the elliptical-beam experiments, the thermal conductivities along the in-plane directions perpendicular to and parallel to the *c*-axis of ZnO are determined to be $K_a$ = 46 W m$^{-1}$ K$^{-1}$ and $K_c$ = 56 W m$^{-1}$ K$^{-1}$, respectively, and the through-plane thermal conductivity is determined as $K_z$ = 46 W m$^{-1}$ K$^{-1}$. The $K_c$ of ZnO determined from the elliptical-beam method (56 W m$^{-1}$ K$^{-1}$) is slightly lower than a first-principles calculation in literature (62 W m$^{-1}$ K$^{-1}$)[35] but is consistent with another TDTR measurement of $K_z$ of ZnO [0001] (55 W m$^{-1}$ K$^{-1}$)[29], which has its *c*-axis along the through-plane direction.

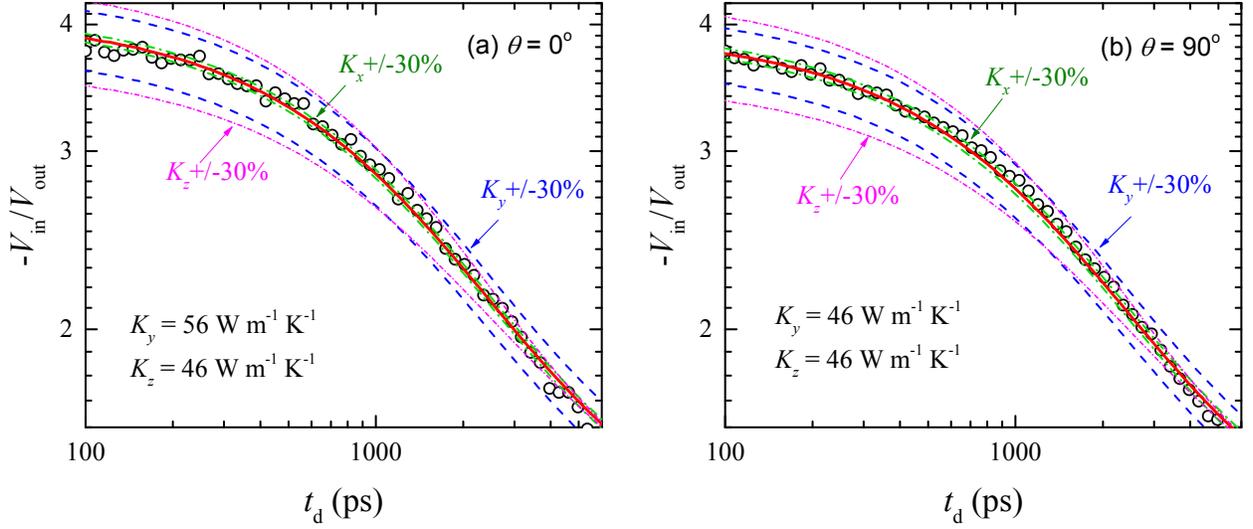

FIG. 9. The ratio signals from the elliptical-beam experiments and their fitting curves for ZnO [11-20] with the short radius of the elliptical beam oriented to be (a) parallel to and (b) perpendicular to the c-axis of ZnO, respectively. The curves of 30% bounds on the best-fitted thermal conductivity values are also included as a guide of reading to the sensitivity of the signals.

Figure 10 shows a summary of the in-plane thermal conductivity tensor of ZnO [11-20] measured by the beam-offset method and the elliptical-beam method under their optimal experimental conditions, respectively. Overall, these two methods compare relatively well with each other, with the measured in-plane thermal conductivities within the error bars. However, the data by the beam-offset method scatter more significantly while the data by the elliptical-beam method show better consistency.

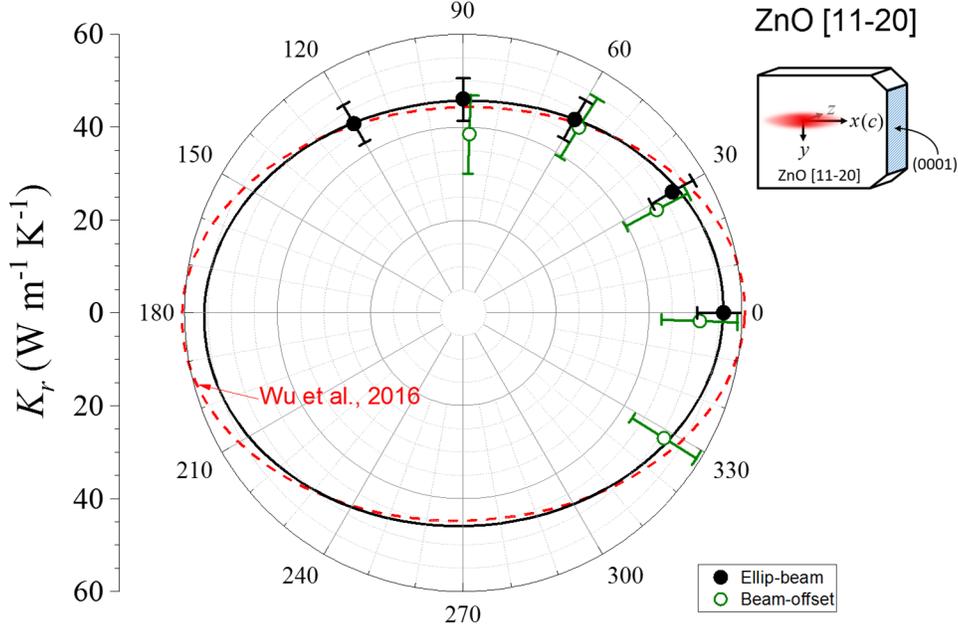

FIG. 10. In-plane thermal conductivity tensor of ZnO [11-20] determined by the elliptical-beam method (solid symbols) and the beam-offset method (open symbols), compared with the first-principles calculations (dashed line) from Ref. 35.

## IV. CONCLUSION

In summary, we have proposed a novel elliptical-beam approach based on TDTR to measure in-plane thermal conductivity tensor of laterally anisotropic materials, and have it compared with the beam-offset approach that was recently proposed in literature[11]. In the elliptical-beam TDTR approach, a highly elliptical pump beam spot is used instead of a circular spot for the TDTR measurements, which suppresses the sensitivity of the TDTR signal to the in-plane thermal conductivity along the direction of major axis of the elliptical beam spot. Through systematic sensitivity analysis, we provide guidelines for the optimal experimental conditions for both methods. The two methods are compared by measuring the in-plane thermal conductivity tensor of a ZnO [11-20] sample under their optimal experimental conditions. The in-plane thermal conductivity tensor measured by the two methods are in agreement with each other, while the elliptical-beam method shows the advantages of better accuracy and smaller measurement uncertainty.

**ACKNOWLEDGMENTS**

This work was supported by NSF Grant No. 1511195 and DOE Grant No. DE-AR0000743.

**APPENDIX A: THERMAL MODEL FOR TDTR EXPERIMENTS ON ANISOTROPIC MATERIALS USING CIRCULAR OR ELLIPTICAL LASER SPOTS WITH OR WITHOUT BEAM OFFSET**

The thermal transport model for the conventional TDTR experiments has been well established.[30,32] Here, we present a generally applicable thermal transport model that applies to the conventional TDTR experiments as well as the beam-offset TDTR experiments and the elliptical-beam TDTR experiments. In this model, when the beam spot sizes $w_x = w_y$ and offset distance $x_c \neq 0$, the model applies to the beam-offset case. When $w_x \neq w_y$ and $x_c = 0$, it applies to the elliptical-beam case. When $w_x = w_y$ and $x_c = 0$, it reduces to the conventional TDTR case.

We first start from the heat diffusion in a multilayered system with anisotropic thermal conductivity in each layer:

$$C\frac{\partial T}{\partial t} = K_x \frac{\partial^2 T}{\partial x^2} + K_y \frac{\partial^2 T}{\partial y^2} + K_z \frac{\partial^2 T}{\partial z^2} + 2K_{xy}\frac{\partial^2 T}{\partial x \partial y} + +2K_{xz}\frac{\partial^2 T}{\partial x \partial z} + +2K_{yz}\frac{\partial^2 T}{\partial y \partial z} \quad (A1)$$

This parabolic partial differential equation can be simplified by doing Fourier transforms with respect to the in-plane coordinates and time, $T(x,y,z,t) \leftrightarrow \Theta(u,v,z,\omega)$, utilizing the following relationships

$$\mathcal{F}\{f(x)\} = F(u) = \int_{-\infty}^{\infty} f(x) e^{-i2\pi u x}\, dx$$

$$\mathcal{F}\left\{\frac{df(x)}{dx}\right\} = i2\pi u F(u)$$

$$\mathcal{F}\left\{\frac{d^2 f(x)}{dx^2}\right\} = -(2\pi u)^2 F(u)$$

as

$$(iC\omega)\Theta = -4\pi^2\left(K_x u^2 + 2K_{xy}uv + K_y v^2\right)\Theta + 2i2\pi\left(K_{xz}u + K_{yz}v\right)\frac{\partial \Theta}{\partial z} + K_z \frac{\partial^2 \Theta}{\partial z^2} \quad (A2)$$

or more compactly,

$$\frac{\partial^2 \Theta}{\partial z^2} + \lambda_2 \frac{\partial \Theta}{\partial z} - \lambda_1 \Theta = 0 \quad (A3)$$

where

$$\lambda_1 \equiv \frac{iC\omega}{K_z} + \frac{4\pi^2(K_x u^2 + 2K_{xy}uv + K_y v^2)}{K_z} \quad (A4)$$

$$\lambda_2 \equiv i4\pi \frac{K_{xz}u + K_{yz}v}{K_z} \quad (A5)$$

The general solution of Eq. (A3) is

$$\Theta = e^{u^+ z} B^+ + e^{u^- z} B^- \quad (A6)$$

where $u^+, u^-$ are the roots of the equation $x^2 + \lambda_2 x - \lambda_1 = 0$:

$$u^\pm = \frac{-\lambda_2 \pm \sqrt{(\lambda_2)^2 + 4\lambda_1}}{2} \quad (A7)$$

and $B^+, B^-$ are the complex numbers to be determined.

From the Fourier's law of heat conduction $Q = -K_z(d\Theta/dz)$ and Eq. (A6), the heat flux can be expressed as:

$$Q = -K_z u^+ e^{u^+ z} B^+ - K_z u^- e^{u^- z} B^- \quad (A8)$$

It is convenient to rewrite Eqs. (A6) and (A8) in matrices as

$$\begin{bmatrix} \Theta \\ Q \end{bmatrix}_{n,z} = [N]_n \begin{bmatrix} B^+ \\ B^- \end{bmatrix}_n \quad (A9)$$

$$[N]_n = \begin{bmatrix} 1 & 1 \\ -K_z u^+ & -K_z u^- \end{bmatrix} \begin{bmatrix} e^{u^+ z} & 0 \\ 0 & e^{u^- z} \end{bmatrix}_n \quad \text{(A10)}$$

where $n$ stands for the $n$-th layer of the multilayer system, and $z$ is the distance within the $n$-th layer from its surface.

The constants $B^+, B^-$ for the $n$-th layer can also be obtained from the surface temperature and heat flux of that layer by setting $z = 0$ in Eq. (A10) and calculating the inverse matrix of Eq. (A9) as:

$$\begin{bmatrix} B^+ \\ B^- \end{bmatrix}_n = [M]_n \begin{bmatrix} \Theta \\ Q \end{bmatrix}_{n,z=0} \quad \text{(A11)}$$

$$[M]_n = \frac{1}{K_z(u^+ - u^-)} \begin{bmatrix} -K_z u^- & -1 \\ K_z u^+ & 1 \end{bmatrix} \quad \text{(A12)}$$

For heat flow across the interface, the heat flux and the temperature can be expressed as

$$Q_{n,z=L} = Q_{n+1,z=0} = G(\Theta_{n,z=L} - \Theta_{n+1,z=0}) \quad \text{(A13)}$$

$$\Theta_{n+1,z=0} = \Theta_{n,z=L} - \frac{1}{G} Q_{n,z=L} \quad \text{(A14)}$$

where $G$ is the interface thermal conductance. It is convenient to rewrite Eqs. (A13) and (A14) in matrices as

$$\begin{bmatrix} \Theta \\ Q \end{bmatrix}_{n+1,z=0} = [R]_n \begin{bmatrix} \Theta \\ Q \end{bmatrix}_{n,z=L} \quad \text{(A15)}$$

$$[R]_n = \begin{bmatrix} 1 & -1/G \\ 0 & 1 \end{bmatrix} \quad \text{(A16)}$$

The temperature and heat flux on the surface of the first layer can thus be related to those at the bottom of the substrate as

$$\begin{bmatrix} \Theta \\ Q \end{bmatrix}_{n,z=L_n} = [N]_n [M]_n \cdots [R]_1 [N]_1 [M]_1 \begin{bmatrix} \Theta \\ Q \end{bmatrix}_{1,z=0} = \begin{bmatrix} A & B \\ C & D \end{bmatrix} \begin{bmatrix} \Theta \\ Q \end{bmatrix}_{1,z=0} \quad \text{(A17)}$$

Applying the boundary condition that the heat flux at the bottom of the substrate is zero, there is $0 = C\Theta_1 + DQ_1$. The temperature response function $H$, which is the detected temperature change in response to the applied heat flux, can thus be found out as

$$H(u,v,\omega) = \frac{\Theta_1}{Q_1} = -\frac{D}{C} \tag{A18}$$

The next step is to simulate the heating and signal detection in TDTR experiments. The sample surface is heated by an elliptical pump beam that has a Gaussian distribution of intensity $p_0(x,y)$ expressed as

$$p_0(x,y) = \frac{2A_0}{\pi \sigma_{x_0} \sigma_{y_0}} \exp\left(-\frac{2x^2}{\sigma_{x_0}^2}\right) \exp\left(-\frac{2y^2}{\sigma_{y_0}^2}\right) \tag{A19}$$

where $\sigma_{x_0}$ and $\sigma_{y_0}$ are the $1/e^2$ radii of the pump spot in the $x$ and $y$ directions respectively. The 2-D Fourier transform of $p_0(x,y)$ utilizing the following relationships

$$\mathcal{F}\{f(x,y)\} = F(u,v) = \iint f(x,y) e^{-i2\pi(ux+vy)} dx dy$$

$$\mathcal{F}\{e^{-ax^2}\} = \int_{-\infty}^{\infty} e^{-ax^2} e^{-i2\pi ux} dx = \sqrt{\left(\frac{\pi}{a}\right)} e^{-\pi^2 u^2/a}$$

yields

$$P_0(u,v) = A_0 \exp\left(-\frac{\pi^2 u^2 \sigma_{x_0}^2}{2}\right) \exp\left(-\frac{\pi^2 v^2 \sigma_{y_0}^2}{2}\right) \tag{A20}$$

The distribution of surface temperature oscillation is the inverse transform of the product of the heat flux $P_0(u,v)$ and the temperature response function $H(u,v)$

$$\theta(x,y) = \int_{-\infty}^{\infty}\int_{-\infty}^{\infty} P_0(u,v) H(u,v) e^{i2\pi(ux+vy)} du dv \tag{A21}$$

The surface temperature oscillation is measured as a weighted average by an elliptical probe beam with $x$- and $y$-offsets to the pump as

$$\Delta T = \frac{2}{\pi \sigma_{x_1} \sigma_{y_1}} \int_{-\infty}^{\infty} \int_{-\infty}^{\infty} \theta(x,y) \exp\left(-\frac{2(x-x_c)^2}{\sigma_{x_1}^2}\right) \exp\left(-\frac{2(y-y_c)^2}{\sigma_{y_1}^2}\right) dxdy \quad (A22)$$

where $\sigma_{x_1}$ and $\sigma_{y_1}$ are the $1/e^2$ radii of the probe spot in the $x$ and $y$ directions, respectively, and $x_c$ and $y_c$ are the offset distance between the pump and the probe in the $x$ and $y$ directions, respectively.

The integral of $\theta$ over $x$ and $y$ in Eq. (A22) is the inverse Fourier transform of the probe beam, leaving an integral over $u$ and $v$ that must be evaluated numerically

$$\Delta T = \int_{-\infty}^{\infty} \int_{-\infty}^{\infty} H(u,v) \exp(-\pi^2 u^2 w_x^2) \exp(-\pi^2 v^2 w_y^2) \exp(i2\pi(ux_c + vy_c)) dudv \quad (A23)$$

where $w_x^2 = (\sigma_{x_0}^2 + \sigma_{x_1}^2)/2$, $w_y^2 = (\sigma_{y_0}^2 + \sigma_{y_1}^2)/2$.

The signal detected by the lock-in amplifier is[37]

$$\Delta R_M(\omega_0) = \frac{dR}{dT} \sum_{n=-\infty}^{\infty} \Delta T(\omega_0 + n\omega_s) e^{in\omega_s t_d} \quad (A24)$$

where $\omega_0$ is the modulation frequency of pump heating, $\omega_s$ is the sampling frequency by the laser pulses (i.e., $2\pi$ times the laser repetition rate), $t_d$ is the delay time between pump and probe, and $dR/dT$ is the thermoreflectance coefficient. More specifically, lock-in amplifier will have in-phase and out-of-phase outputs which are the real and imaginary parts of $\Delta R_M$ respectively:

$$V_{in} = \text{Re}\{\Delta R_M(\omega_0)\} = \frac{1}{2}\frac{dR}{dT} \sum_{n=-\infty}^{\infty} [\Delta T(\omega_0 + n\omega_s) + \Delta T(-\omega_0 + n\omega_s)] e^{in\omega_s t_d} \quad (A25)$$

$$V_{out} = \text{Im}\{\Delta R_M(\omega_0)\} = -\frac{i}{2}\frac{dR}{dT} \sum_{n=-\infty}^{\infty} [\Delta T(\omega_0 + n\omega_s) - \Delta T(-\omega_0 + n\omega_s)] e^{in\omega_s t_d} \quad (A26)$$

The ratio $R = -V_{in}/V_{out}$ a function of delay time $t_d$ is usually taken as the signal to extract the unknown thermal properties by comparing the thermal model calculations to the measurements.

**APPENDIX B: SOURCES OF SENSITIVITY TO $K_x$ AND $K_z$ IN TDTR EXPERIMENTS**

Several questions are raised about the elliptical-beam method: Why are the sensitivities to $K_x$ and $K_z$ not correlated to each other, as depicted in Figure 3(a)? Does the anisotropic thermal conductivity ratio $K_x/K_z$ affect the measurements of in-plane thermal conductivity? Is it possible that the ratio signal in TDTR experiments is only sensitive to the in-plane thermal conductivity $K_x$ but not to the through-plane thermal conductivity $K_z$? To answer these questions, we need to have a deeper understanding on the sources of sensitivity in TDTR experiments.

In TDTR experiments, the signals have an in-phase part $V_{in}$ and an out-of-phase part $V_{out}$. The in-phase signal $V_{in}$ further contains two components: one is due to the single pulse heating, $\Delta V_{in}$, and the other is due to the pulse accumulation, $V_{in} = \Delta V_{in} + V_{in}(t_d < 0)$. Here we take the sample of 100 nm Al/Si as an example to discuss how the different components of the TDTR signals ($\Delta V_{in}$, $V_{in}$, and $-V_{out}$) are sensitive to the thermal properties of the sample (in-plane and through-plane thermal conductivity of the substrate, $K_x$ and $K_z$, and the Al/substrate interface thermal conductance $G$) under different configurations of laser spot ($w_x$ = 1 μm and 50 μm) and modulation frequency ($f$ = 0.1 MHz and 10 MHz). The sensitivity coefficients of the signals $\Delta V_{in}$, $V_{in}$, and $-V_{out}$ are defined in the same way as in Eq. (1) in the main text, only to have the ratio signal $R$ in Eq. (1) replaced with the corresponding signals. In this calculation, the thermal properties are assumed to be $K_m$ = 180 W m$^{-1}$ K$^{-1}$ and $C_m$ = 2.44 J cm$^{-3}$ K$^{-1}$ for the Al transducer, $G$ = 100 MW m$^{-2}$ K$^{-1}$ for the interface conductance, and $K_x = K_y = K_z$ = 140 W m$^{-1}$ K$^{-1}$ and $C$ = 1.6 J cm$^{-3}$ K$^{-1}$ for the Si substrate. The calculated sensitivity coefficients are summarized in Figure B1. Here we take the x-direction as the representative one to discuss the in-plane heat conduction, while the discussion here applies to other in-plane directions as well. Several general conclusions on the sources of measurement sensitivity can be drawn from the sensitivity plots in Figure B1 and are summarized below.

(1) The in-phase signal $\Delta V_{in}$ is the result of single pulse heating; therefore, it is independent of the modulation frequency. The signal $\Delta V_{in}$ becomes sensitive to the interface conductance $G$ after the pulsed heating has diffused across the Al transducer films (the diffusion time can be estimated as $\tau = C_{Al} h_{Al}^2 / K_{Al}$ and is ~135 ps for 100 nm Al transducer). The signal $\Delta V_{in}$ also becomes sensitive to the through-plane thermal conductivity of the substrate $K_z$ after the pulsed heating has diffused across the Al/substrate interface (the diffusion time can be estimated as $\tau = C_{Al} h_{Al} / G$ and is ~2.4 ns for a $G$ of 100 MW m$^{-2}$ K$^{-1}$). The sensitivities of $\Delta V_{in}$ to the through-plane properties $K_z$ and $G$ are independent of the laser spot size (1 – 50 μm), suggesting that the pulsed heating diffuses in the through-plane direction irrespective of the laser spot size. For most cases, the signal $\Delta V_{in}$ is not sensitive to the in-plane thermal conductivity of the substrate $K_x$ due to the much shorter thermal diffusion length $d_f = \sqrt{K\tau/C}$ compared to the laser spot size. Here the time scale $\tau$ is limited by the maximum delay time of 12.5 ns. However, when the laser spot size is comparable to $d_f$, the signal $\Delta V_{in}$ also becomes sensitive to $K_x$. For example, for Si substrate at delay time 10 ns, the heat diffusion length is estimated to be $d_f \approx 1$ μm, similar to the laser spot sizes in cases (a1) and (b1) in Figure B1, thus the signal $\Delta V_{in}$ is also sensitive to $K_x$ at delay time 10 ns in cases (a1) and (b1).

(2) When the time interval between pulses (12.5 ns) is not long enough for the in-phase temperature rise to fully decay to its original state, there would be pulse accumulation. Of the four cases in Figure B1, only case (c) has strong pulse accumulation, with the magnitude of the accumulated in-phase signal $V_{in}(t_d < 0)$ being similar to or larger than that of the in-phase jump $\Delta V_{in}$, as shown in Figure B2. The effect of strong pulse accumulation makes the $V_{in}$ signal more sensitive to $K_z$ of the substrate and less sensitive to the interface conductance $G$, as revealed by comparing case (c2) to others.

(3) Whether the out-of-phase signal $-V_{out}$ is sensitive to the in-plane thermal conductivity $K_x$ or not depends on how the in-plane thermal penetration depth $d_{p,x}$ is compared to the laser spot size $w_x$. When $d_{p,x} \gg w_x$, the $-V_{out}$ signal is highly sensitive to $K_x$, and vice versa. For the cases in Figure B1, the in-plane thermal penetration depths $d_{p,x}$ are ~17 μm at 0.1 MHz and ~1.7 μm at 10 MHz. The $-V_{out}$ signal is highly sensitive to $K_x$ in case (a3) because $d_{p,x} \gg w_x$. The $-V_{out}$ signal is not sensitive to $K_x$ in case (d3) because $d_{p,x} \ll w_x$. The $-V_{out}$ signal is moderately sensitive to $K_x$ in cases (b3) and (c3) because $d_{p,x} \approx w_x$.

(4) Whether the out-of-phase signal $-V_{out}$ is sensitive to $K_z$ and $G$ or not depends on how the through-plane thermal penetration depth $d_{p,z}$ is compared to the equivalent thickness of the interface, defined as $h_G = K_z/G$. When $d_{p,z} \gg h_G$, the $-V_{out}$ signal is sensitive to $K_z$ but not $G$. When $d_{p,z} \ll h_G$, the $-V_{out}$ signal becomes sensitive to $G$ but not $K_z$. When $d_{p,z} \approx h_G$, the $-V_{out}$ signal becomes sensitive to both $K_z$ and $G$. For the cases in Figure B1, the through-plane thermal penetration depths $d_{p,z}$ are ~17 μm at 0.1 MHz and ~1.7 μm at 10 MHz, and the equivalent interface thickness is $h_G = 1.4$ μm. The $-V_{out}$ signal is highly sensitive to $K_z$ but not $G$ in cases (a3) and (c3) because $d_{p,z} \gg h_G$. The $-V_{out}$ signal is sensitive to both $K_z$ and $G$ in cases (b3) and (d3) because $d_{p,z} \approx h_G$. The sensitivities of the $-V_{out}$ signal to the through-plane thermal properties $K_z$ and $G$ are little affected by the laser spot size (1 – 50 μm), suggesting that the continuous heating diffuses in the through-plane direction irrespective of the laser spot size.

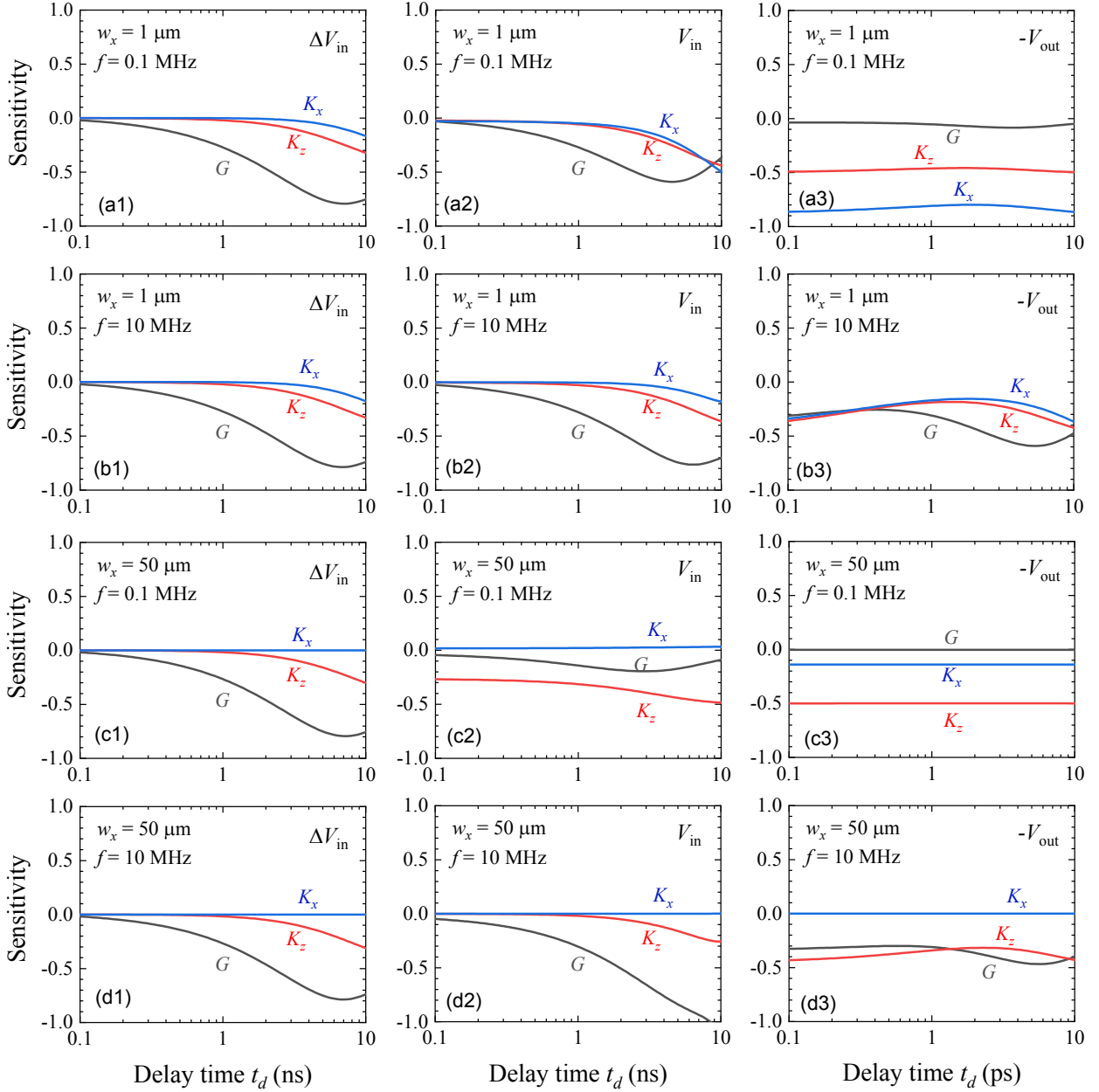

FIG. B1. Sensitivity coefficients of the signals $\Delta V_{in}$, $V_{in}$, and $-V_{out}$ to different parameters ($K_x$, $K_z$, and $G$) of the sample 100 nm Al/Si as a function of the delay time under different experimental conditions of laser spot sizes and the modulation frequency: (a1-a3) $w = 1\ \mu m$, $f = 0.1$ MHz; (b1-b3) $w = 1\ \mu m$, $f = 10$ MHz; (c1-c3) $w = 50\ \mu m$, $f = 0.1$ MHz; (d1-d3) $w = 50\ \mu m$, $f = 10$ MHz.

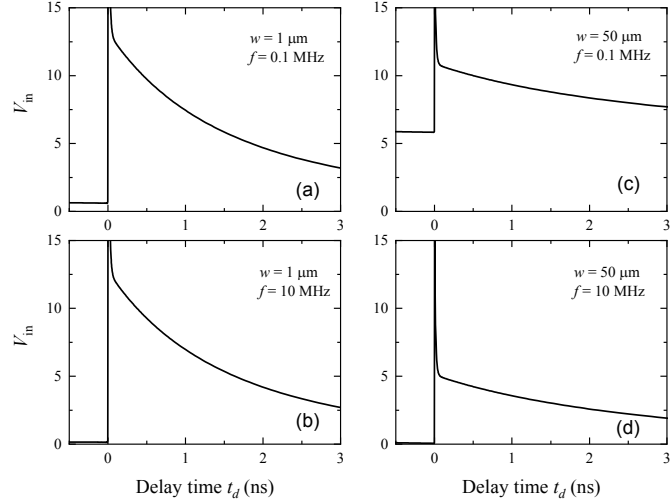

FIG. B2. Simulated in-phase signals as a function of the delay time for the four cases in Figure B1. Only case (c) has the effect of significant pulse accumulation.

From the discussion above, we understand that the sensitivities of the TDTR signals to $K_x$ and $K_z$ come from different sources and do not affect each other. While the TDTR signals (both $V_{in}$ and $V_{out}$) are always sensitive to through-plane thermal properties ($G$ and/or $K_z$) because of the inevitable through-plane heat flow, the TDTR signals (mainly $V_{out}$) become sensitive to in-plane thermal properties ($K_x$) only when the laser spot size $w_x$ is comparable or smaller than the in-plane thermal penetration depth $d_{p,x}$ in the same in-plane direction.

When the ratio $R = -V_{in}/V_{out}$ is taken as the TDTR signal to derive thermal properties, the sensitivity coefficient of the ratio signal $R$ can be viewed as the difference of the sensitivity coefficients of its two components:

$$S_\alpha^R = \frac{\partial \ln(-V_{in}/V_{out})}{\partial \ln \alpha} = \frac{\partial \ln(V_{in})}{\partial \ln \alpha} - \frac{\partial \ln(-V_{out})}{\partial \ln \alpha} = S_\alpha^{V_{in}} - S_\alpha^{-V_{out}} \tag{B1}$$

For most cases, the in-phase signal $V_{in}$ is initially not sensitive to $K_z$ of the substrate at a short delay time of 100 ps, and it starts to become more sensitive to $K_z$ at the longer delay times, while the out-of-phase signal -$V_{out}$ is constantly sensitive to $K_z$ over the whole delay time range. As a result, the ratio signal $R = -V_{in}/V_{out}$ is initially sensitive to $K_z$ at the short delay time (the sensitivity comes mainly from $V_{out}$) and the sensitivity diminishes at longer delay times (the sensitivities from $V_{in}$ and $V_{out}$ are canceled due to similar amplitudes). On the other hand, usually only the out-of-phase signal -$V_{out}$ is sensitive to $K_x$ but the in-phase signal $V_{in}$ is not. As a result, the ratio signal $R = -V_{in}/V_{out}$ is constantly sensitive to $K_x$ over the whole delay time range. This explains why the sensitivities of the ratio signal $R$ to the in-plane and through-plane thermal conductivities of the substrate are not correlated over the delay time, as depicted in Figure 3(a) in the main text.